# QUANTUM SYSTEM WITH A SHOCK WAVE

## E.E. Perepelkin[a] B.I. Sadovnikov[a], N.G. Inozemtseva[b]

[a] Faculty of Physics, Lomonosov Moscow State University, Moscow, 119991 Russia
e-mail: pevgeny@jinr.ru
[b] Dubna University, Moscow region, Moscow,141980 Russia
e-mail: nginozv@mail.ru

**Abstract**
The experimental shock wave, discovery in quantum systems is the important one through last years. This effect has several phenomenological explanations. It seems to us there is interesting, especially from the systematical point of view, for high school student and scientists alike, to take the result from «first principles», that is «with the tip of the pen». It was made in the paper proposed.

## Introduction

In paper [1] we have determined a strict interchangeability between the continuum equation (first equation of the Vlasov chain of equations [2-4]) for the probability density function $f(\vec{r},t) = \Psi(\vec{r},t)\bar{\Psi}(\vec{r},t) = |\Psi(\vec{r},t)|^2$ and the Schrödinger equation. The following statement for the velocity of the probability flow $\langle\vec{v}\rangle(\vec{r},t)$ according to the Helmholtz decomposition has been obtained [1]:

$$\langle\vec{v}\rangle(\vec{r},t) = i\alpha\nabla\mathrm{Ln}\left[\frac{\Psi}{\bar{\Psi}}\right] + \gamma\vec{A}, \qquad (i.1)$$

where $\nabla\mathrm{Ln}\left[\frac{\Psi}{\bar{\Psi}}\right] = \nabla\Phi(\vec{r},t)$ is corresponds with the irrotational one, and $\vec{A}$ is with the solenoidal component of the velocity of the probability flow. The constants $\alpha$ and $\gamma$ are equal to $-\frac{\hbar}{2m}$ and $-\frac{e}{m}$ in case of considering the quantum system. In this way, the continuum equation for the probability density function $f(\vec{r},t)$:

$$\frac{\partial f(\vec{r},t)}{\partial t} + \mathrm{div}_r\left[f(\vec{r},t)\langle\vec{v}\rangle(\vec{r},t)\right] = 0, \qquad (i.2)$$

became the equation [1, 24, 25]:

$$\frac{i}{\beta}\frac{\partial\Psi}{\partial t} = -\alpha\beta\left(\hat{\mathrm{p}} - \frac{\gamma}{2\alpha\beta}\vec{A}\right)^2\Psi + \frac{1}{2\alpha\beta}\frac{|\gamma\vec{A}|^2}{2}\Psi + U\Psi \qquad (i.3)$$

where $\hat{\mathrm{p}} = -\frac{i}{\beta}\nabla$, $\beta = \frac{1}{\hbar}$, and the potential $U$ has the following form:

$$U(\vec{r},t) = -\frac{1}{\beta}\left\{\frac{\partial \varphi(\vec{r},t)}{\partial t} + \alpha\left[\frac{\Delta\sqrt{f(\vec{r},t)}}{\sqrt{f(\vec{r},t)}} - |\nabla\varphi(\vec{r},t)|^2\right] + \gamma(\vec{A},\nabla\varphi)\right\}, \qquad (i.4)$$

where $\varphi(\vec{r},t)$ is the phase of the wave function $\Psi(\vec{r},t)$, which is directly connected the scalar potential $\Phi(\vec{r},t)$ of the velocity of the probability flow (i.1), as

$$\mathrm{Arg}\left[\frac{\Psi(\vec{r},t)}{\overline{\Psi}(\vec{r},t)}\right] = 2\varphi(\vec{r},t) + 2\pi k = \Phi(\vec{r},t), \qquad (i.5)$$

as

$$\mathrm{Ln}\left[\frac{\Psi}{\overline{\Psi}}\right] = \ln\left|\frac{\Psi}{\overline{\Psi}}\right| + i\,\mathrm{Arg}\left[\frac{\Psi}{\overline{\Psi}}\right] = i\Phi(\vec{r},t). \qquad (i.6)$$

In [1, 24, 25] the following equation was obtained:

$$\frac{\partial \Phi}{\partial t} = -\frac{2}{\hbar}\left[\frac{m|\langle\vec{v}\rangle|^2}{2} + e\chi\right] = -\frac{2}{\hbar}W(\vec{r},t), \qquad (i.7)$$

where $e\chi$ is the potential energy, $T = \dfrac{m|\langle\vec{v}\rangle|^2}{2}$ is kinetic, and $W(\vec{r},t)$ is total energy of the system. The potential $U$ (i.4) is associated with the classical potential $e\chi$ (i.7) with ratio

$$\chi \stackrel{\mathrm{det}}{=} \frac{2\alpha\beta}{\gamma}\left(\frac{1}{2\alpha\beta}\frac{|\gamma\vec{A}|^2}{2} + U + \frac{\alpha}{\beta}\frac{\Delta\sqrt{f}}{\sqrt{f}}\right). \qquad (i.8)$$

Note that expression (i.8) contains a summand $\dfrac{\alpha}{\beta}\dfrac{\Delta\sqrt{f}}{\sqrt{f}} = -\dfrac{\hbar^2}{2m}\dfrac{\Delta\sqrt{f}}{\sqrt{f}}$ which is a well-known quantum potential in the theory of pilot wave [13-15, 24, 25]. Thus in (i.8) we can find the term $\dfrac{|\gamma\vec{A}|^2}{2}$ corresponding to the kinetic energy of the solenoidal velocity field. The potentials (i.4) and (i.8) are obtained in [1] based only on equation (i.2) and representation (i.3).

As the solutions of equation (i.2) are strongly connected with the solutions of the Schrödinger equation (i.3), we can obtain similar solutions of the Schrödinger equation (i.3) for the case of the irrotational flow probability according to the hydrodynamic solutions of the equation (i.2). A solution of the shock wave is among such solutions [16-23]. This paper is devoted to obtaining these solutions and components corresponding with them.

The Sect. 1 contains the information about the obtaining of the exact solutions of the equation (i.2) by the method of characteristics for two types of symmetries of the initial distributions of the density $f(\vec{r},t)$ [5]. The spherically and cylindrically symmetric distributions $f(\vec{r},t)$ are also considered in this section. In addition, there is a consideration of two types of interactions – the electric one and the gravitational. The characteristics equations, which are the classical mechanical trajectories of the concentric layers of the charge and mass distribution, are obtained. Intersection of the characteristics leads to the solutions of the shock wave type. It is noteworthy to say that in case of electric interaction solutions of this type are significant in the Coulomb explosion study [6-11].

In Sect. 2 we have obtained the potential of interaction $U(\vec{r},t)$ of the Schrödinger equation (i.3) (at $\alpha = -\dfrac{\hbar}{2m}$, $\beta = \dfrac{1}{\hbar}$, $\vec{A} = \vec{0}$), and the wave function $\Psi(\vec{r},t)$ and its phase $\varphi(\vec{r},t)$ which, according to (i.1) and (i.5), within the accuracy of the coefficient is a scalar potential $\Phi(\vec{r},t)$ of the velocity of the probability flow $\langle\vec{v}\rangle(\vec{r},t)$.

The Sect. 3 contains the consideration of the probability distribution function formation. It is shown that according to the geometric point of view the surface of the probability distribution function consists of the characteristics obtained in the Sect. 1 and which are the classical mechanical trajectories [12]. The descriptive geometric interpretation of the connection between the probabilistic (quantum) and deterministic (classical) description of the system is given.

### 1. Hydrodynamic solutions

Let us apply the method of characteristics and make the exact solutions of the evolution problem of the charge or mass density function for areas having the form of a sphere or an infinite cylinder. The initial distribution of the charge or mass density for the area of a sphere type has the spherical symmetry, i.e. there is the dependence on the radius $r$. For the area of an infinite cylindrical type the initial distribution of the charge or mass density has the azimuth and longitudinal symmetry, that is why there is the dependence only on the radius $r$.

### 1.1 Electric interaction

Let us consider the system consisting of electric charges $q$ with the mass $m$. Let the initial density of the charge distribution $\rho_0(r)$ of such system be given. The initial velocity of the particles in the system is considered to be equal to zero. We are going to solve the evolution problem of the density function of the charge $\rho_s(r,t)$ ( or $\rho_c(r,t)$ ) of the given system. Index «s» («c») agrees with the spherically (cylindrical) symmetric case.

Owing to the symmetry of the considered system the charged particles are under the influence of the Coulomb repulsion. During the solution we are going to suppose that the charged spherical layers will not cross each other in the course of time. According to the Gauss theorem, the electric field on the surface of the sphere with the radius $r$ for any radius $r$ is presented as:

$$\vec{D}_s(r,t) = \vec{e}_r \frac{1}{r^2} \int_0^r \rho_s(x,t) x^2 dx, \quad \vec{D}_c(r,t) = \vec{e}_r \frac{1}{r} \int_0^r \rho_c(x,t) x dx. \tag{1}$$

Let us write down the Newton's second law for the charge which is on the surface of the sphere (cylinder) with the radius $R$:

$$R_{tt} = \frac{q}{m\varepsilon_0} D_s(R,t) = \frac{q}{m\varepsilon_0} \frac{1}{R^2} \int_0^R \rho_s(x,t) x^2 dx,$$

$$R_{tt} = \frac{q}{m\varepsilon_0} D_c(R,t) = \frac{q}{m\varepsilon_0} \frac{1}{R} \int_0^R \rho_c(x,t) x dx, \tag{2}$$

where $\varepsilon_0$ is the permittivity. As the spherical (cylindrical) layers do not cross each other the charge which is in the sphere (cylinder) with the radius $R_0 = R(t=0)$ stays invariable during the whole movement. The quantity of this charge may be defined by the following formula:

$$Q_s(R_0) = Q_s(R(t)) = 4\pi \int_0^{R_0} \rho_s(x,0) x^2 dx = 4\pi \int_0^{R(t)} \rho_s(x,t) x^2 dx,$$

$$Q_c(R_0) = Q_c(R(t)) = 2\pi \int_0^{R_0} \rho_c(x,0) x dx = 2\pi \int_0^{R(t)} \rho_c(x,t) x dx. \qquad (3)$$

Taking (3) into account, expressions (2) will have the following form:

$$R_{tt} = \frac{q}{m\varepsilon_0} \frac{Q_s(R_0)}{4\pi R^2}, \qquad R_{tt} = \frac{q}{m\varepsilon_0} \frac{Q_c(R_0)}{2\pi R}. \qquad (4)$$

As a result we obtain the Cauchy problem for the function $R(t)$:

$$\begin{cases} R'' = \gamma_s \dfrac{1}{R^2}, \\ R\big|_{t=0} = R_0, \\ R'\big|_{t=0} = 0, \end{cases} \qquad \begin{cases} R'' = \gamma_c \dfrac{1}{R}, \\ R\big|_{t=0} = R_0, \\ R'\big|_{t=0} = 0, \end{cases} \qquad (5)$$

where $\gamma_s = \dfrac{q}{4\pi\varepsilon_0 m} Q_s(R_0)$, $\gamma_c = \dfrac{q}{2\pi\varepsilon_0 m} Q_c(R_0)$ are the constant values.

*Solution*

Let us determine the solution for (5) by multiplying both parts of the differential equation from (5) by $R'$, and then integrating with $t$:

$$(R')^2 = C_1^s - \gamma_s \frac{2}{R},$$
$$(R')^2 = 2\gamma_c \ln R + C_1^c, \qquad (6)$$

where $C_1$ is a constant value. Taking the initial condition $R'(0) = 0$ into account, we obtain $C_1^s = \gamma_s \dfrac{2}{R_0}$, $C_1^c = -2\gamma_c \ln R_0$. As a result the expression (6) takes the form of:

$$\frac{1}{\sqrt{2\gamma_s}} \frac{dR}{\sqrt{\dfrac{1}{R_0} - \dfrac{1}{R}}} = dt, \qquad \frac{1}{\sqrt{2\gamma_c}} \frac{dR}{\sqrt{\ln \dfrac{R}{R_0}}} = dt. \qquad (7)$$

Integrating with (7) and taking the initial condition $R(0) = R_0$ into account, we obtain the expressions determining the dependence of the spherical (cylindrical) layer radius on time

$$t = \frac{R_0^{3/2}}{\sqrt{2\gamma_s(R_0)}} \left( \sqrt{\frac{R}{R_0}\left(\frac{R}{R_0}-1\right)} + arcch\sqrt{\frac{R}{R_0}} \right),$$

$$\frac{\sqrt{2\gamma_c}}{2R_0} t = \int_0^{\sqrt{\ln(R/R_0)}} e^{l^2} dl. \qquad (8)$$

On the other hand, expressions (8) may be considered as the characteristics equations corresponding with different initial conditions $R_0$, $\gamma(R_0)$. At that, the obtained solution makes sense if the characteristics do not cross each other, ant that corresponds with the non-crossing spherical (cylindrical) layers. Each spherical (cylindrical) layer is characterized by the start position of $R_0$, quantity of the charge $Q_s(R_0)$, $Q_c(R_0)$ inside the sphere (cylinder) with the radius $R_0$, and by the characteristic (2) describing its movement in the course of time.

Let us introduce the symbols:

$$F_s(x) = \sqrt{x(x-1)} + arcch\sqrt{x}, \qquad F_c(x) = \int_0^{\sqrt{\ln(x)}} e^{l^2} dl, \qquad (9)$$

$$\lambda_s(R_0) = \frac{\sqrt{2\gamma_s(R_0)}}{R_0^{3/2}}, \qquad \lambda_c(R) = \frac{\sqrt{2\gamma_c(R)}}{2R}.$$

If the function which is inverse to the function $F(x)$ is defined as $P(F) = x$, then, considering (9), the expressions (8) will take the following form:

$$R(t) = R_0 P(\lambda(R_0)t). \qquad (10)$$

In order to determine the evolution of the density function we are going to evaluate the value $\frac{\Delta Q}{\Delta R}$. $\Delta Q$ is the charge between two spherical (cylindrical) layers with radius $r$ and $r + \Delta r$ respectively. In the course of time the layer with the initial radius $r$ will become the layer with the radius $R_1(t) = rP(\lambda(r)t)$, and the layer with the radius $r + \Delta r$ will become the layer with the radius $R_2(t) = (r + \Delta r)P(\lambda(r + \Delta r)t)$ corresponding with (10). Owing to the fact that the layers do not cross each other, the value $\Delta Q$ to be constant in the course of time, therefore:

$$\Delta Q_s = 4\pi \int_r^{r+\Delta r} x^2 \rho_0(x) dx = 4\pi \int_{R_1(t)}^{R_2(t)} x^2 \rho_s(x,t) dx = 4\pi \rho_0(r) r^2 \Delta r + O(\Delta r^2),$$

$$\Delta Q_c = 2\pi \int_r^{r+\Delta r} x\rho_0(x) dx = 2\pi \int_{R_1(t)}^{R_2(t)} x\rho_c(x,t) dx = 2\pi \rho_0(r) r \Delta r + O(\Delta r^2). \qquad (11)$$

The following correlation is correct for the value $\Delta R$:

$$\Delta R(t) = R_2(t) - R_1(t) = (r + \Delta r) P(\lambda(r + \Delta r)t) - rP(\lambda_s(r)t) =$$
$$= \Delta r \frac{d}{dx}\left(xP(\lambda(x)t)\right)\bigg|_{x=r} + O(\Delta r^2). \quad (12)$$

Substituting (11) and (12) into the correlation $\dfrac{\Delta Q}{\Delta R}$ and considering (3), we obtain:

$$\lim_{\Delta R \to 0} \frac{\Delta Q_s}{\Delta R} = \frac{dQ_s}{dR} = \frac{4\pi \rho_0(r) r^2}{P_s(\lambda_s(r)t) + rP_s'(\lambda_s(r)t)\lambda_s'(r)t} = 4\pi \rho_s(R,t) R^2(t),$$

$$\lim_{\Delta R \to 0} \frac{\Delta Q_c}{\Delta R} = \frac{dQ_c}{dR} = \frac{2\pi \rho_0(r) r}{P_c(\lambda_c(r)t) + rP_c'(\lambda_c(r)t)\lambda_c'(r)t} = 2\pi \rho_c(R,t) R(t), \quad (13)$$

or

$$\rho_s(R,t) = \frac{1}{P_s^2(\lambda_s(r)t)} \frac{\rho_0(r)}{\left[P_s(\lambda_s(r)t) + rP_s'(\lambda_s(r)t)\lambda_s'(r)t\right]}, \quad (14)$$

$$\rho_c(R,t) = \frac{1}{P_c(\lambda_c(r)t)} \frac{\rho_0(r)}{\left[P_c(\lambda_c(r)t) + rP_c'(\lambda_c(r)t)\lambda_c'(r)t\right]},$$

where $R(r,t) = rP(\lambda(r)t)$ in accordance with (10). Let us determine the derivatives $P'(F)$ and $\lambda'(r)$.

$$P(F) = x \Rightarrow P'(F) F'(x) = 1 \Rightarrow P'(F) = \frac{1}{F'(x)}, \quad (15)$$

taking (9) into account we determine $F'(x)$:

$$F_s'(x) = \frac{2x-1}{2\sqrt{x(x-1)}} + \frac{1}{2\sqrt{x-1}\sqrt{x}} = \frac{x}{\sqrt{x(x-1)}} = \frac{P_s}{\sqrt{P_s(P_s-1)}},$$

$$F_c'(x) = e^{(\sqrt{\ln x})^2} \left(\sqrt{\ln x}\right)' = \frac{1}{2\sqrt{\ln x}} = \frac{1}{2\sqrt{\ln P_c}}. \quad (16)$$

As a result, the derivative $P'(F)$ has the following form:

$$P_s'(F_s) = \sqrt{\frac{P_s(F_s) - 1}{P_s(F_s)}}, \qquad P_c'(F_c) = 2\sqrt{\ln P_c(F_c)}. \quad (17)$$

Taking (9) into account for $\lambda(r)$ we determine the derivative $\lambda'(r)$

$$\lambda_s'(r) = \frac{\gamma_s'(r)}{\sqrt{2\gamma_s(r)}} \frac{1}{r^{3/2}} - \frac{3}{2} \frac{\sqrt{2\gamma_s(r)}}{r^{5/2}},$$

$$\lambda_c'(r) = -\frac{\sqrt{2\gamma_c}}{2r^2} + \frac{\gamma_c'}{2r\sqrt{2\gamma_c}}. \tag{18}$$

As the derivative $\gamma'(r)$ has the form

$$\gamma_s'(r) = \delta \frac{r^2}{\varepsilon_0} \rho_0(r), \qquad \gamma_c'(r) = \delta \frac{r}{\varepsilon_0} \rho_0(r), \tag{19}$$

where $\delta = \frac{q}{m}$, then we obtain the following expression for $\lambda'(r)$:

$$\lambda_s'(r) = \frac{\delta r^2 \rho_0(r)}{\varepsilon_0 \sqrt{2\gamma_s(r)}} \frac{1}{r^{3/2}} - \frac{3}{2} \frac{\sqrt{2\gamma_s(r)}}{r^{5/2}} = \frac{1}{r}\left[\frac{\delta \rho_0(r)}{\varepsilon_0 \lambda_s(r)} - \frac{3}{2}\lambda_s(r)\right],$$

$$\lambda_c'(r) = \frac{\delta r \rho_0(r)}{\varepsilon_0 \sqrt{2\gamma_c(r)}} \frac{1}{2r^2} - \frac{\sqrt{2\gamma_c(r)}}{2r} = \frac{1}{r}\left[\frac{\delta \rho_0(r)}{4\varepsilon_0 \lambda_c(r)} - \lambda_c(r)\right]. \tag{20}$$

Substituting expressions (17) and (20) into (14) we obtain the final formula for the charge density:

$$\rho_s(R,t) = \frac{1}{P_s^2(\lambda_s(r)t)} \frac{\rho_0(r)}{\left[P_s(\lambda_s(r)t) + t\sqrt{\frac{P_s(\lambda_s(r)t)-1}{P_s(\lambda_s(r)t)}}\left[\frac{\delta \rho_0(r)}{\varepsilon_0 \lambda_s(r)} - \frac{3}{2}\lambda_s(r)\right]\right]}, \tag{21}$$

$$\rho_c(R,t) = \frac{1}{P_c(\lambda_c(r)t)} \frac{\rho_0(r)}{\left[P_c(\lambda_c(r)t) + 2t\sqrt{\ln P_c(\lambda_c(r)t)}\left[\frac{\delta \rho_0(r)}{4\varepsilon_0 \lambda_c(r)} - \lambda_c(r)\right]\right]},$$

where $\rho(R,t) = \rho(rP(\lambda(r)t),t)$. The formula (21) determines the evolution of the charge density function with the initial distribution $\rho_0(r)$ at the condition that the spherical (cylindrical) layers do not cross each other.

Let us determine the spherical (cylindrical) wave velocity. Let us differentiate the expression (10) with time:

$$\frac{d}{dt} R_s(t) = V_s(t) = R_0 \lambda_s \sqrt{\frac{P_s(\lambda_s t)-1}{P_s(\lambda_s t)}},$$

$$\frac{d}{dt} R_c(t) = V_c(t) = R_0 \lambda_c 2\sqrt{\ln P_c(\lambda_c t)}. \tag{22}$$

Let us determine the maximum front wave velocity. Initial distributions of the charge density not leading to the crossing of the characteristics, for instance, a uniformly charged sphere (cylinder), are of a special interest. In this case the obtained expression (22) is correct everywhere. The function $P(x)$ is uniformly increasing. The minimum value of the function $P(x)$ is 1 at point 0, i. e. $P(0)=1$. For that reason we are going to consider the limit from the expression (22) at $t \to +\infty$.

$$V_{\max}^s = \lim_{t \to +\infty} V_s(t) = R_0 \lambda_s \lim_{t \to +\infty} \sqrt{\frac{P_s(\lambda_s t) - 1}{P_s(\lambda_s t)}} = R_0 \lambda_s \lim_{t \to +\infty} \sqrt{1 - \frac{1}{P(\lambda_s t)}} = R_0 \lambda_s,$$

$$V_{\max}^c = \lim_{t \to +\infty} V_c(t) = 2R_0 \lambda_c \lim_{t \to +\infty} \sqrt{\ln P_c(\lambda_c t)} = +\infty. \qquad (23)$$

Fig. 1 and 2 illustrate the dependence of the velocity on time for various spherical and cylindrical layers respectively.

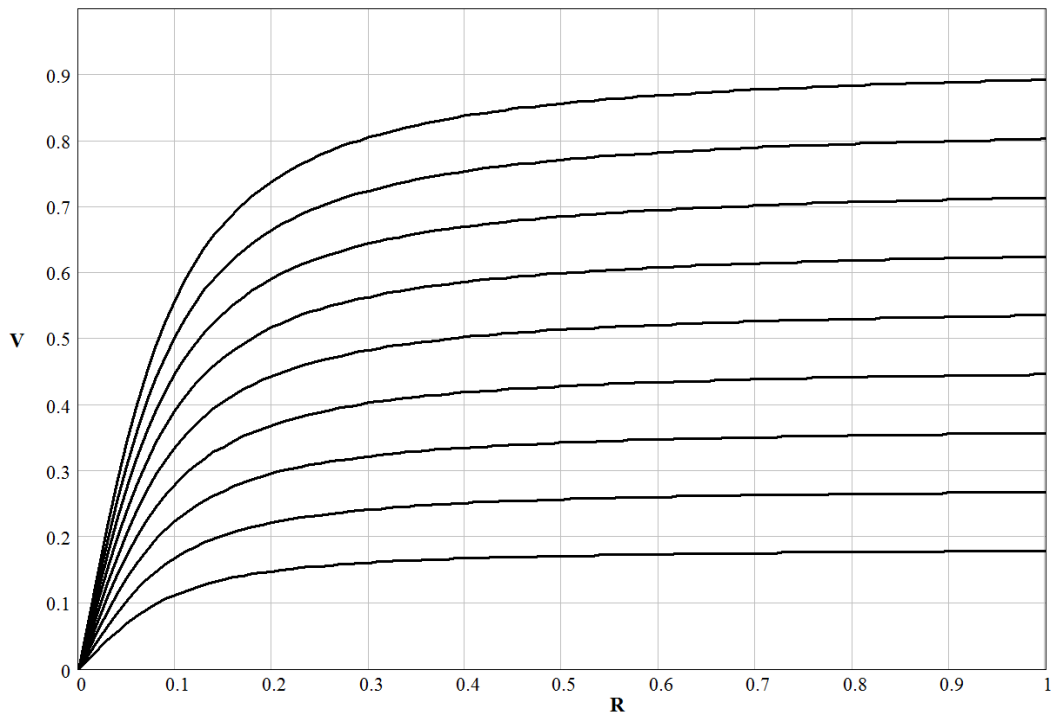

Fig. 1 Dependence of the spherical layers velocity on time for the uniformly charged sphere

According to (23), there is no limit of the maximum velocity for the cylindrical wave $V_{\max}^c$. There is the limit of the maximum front velocity of the spherical wave $V_{\max}^s$.

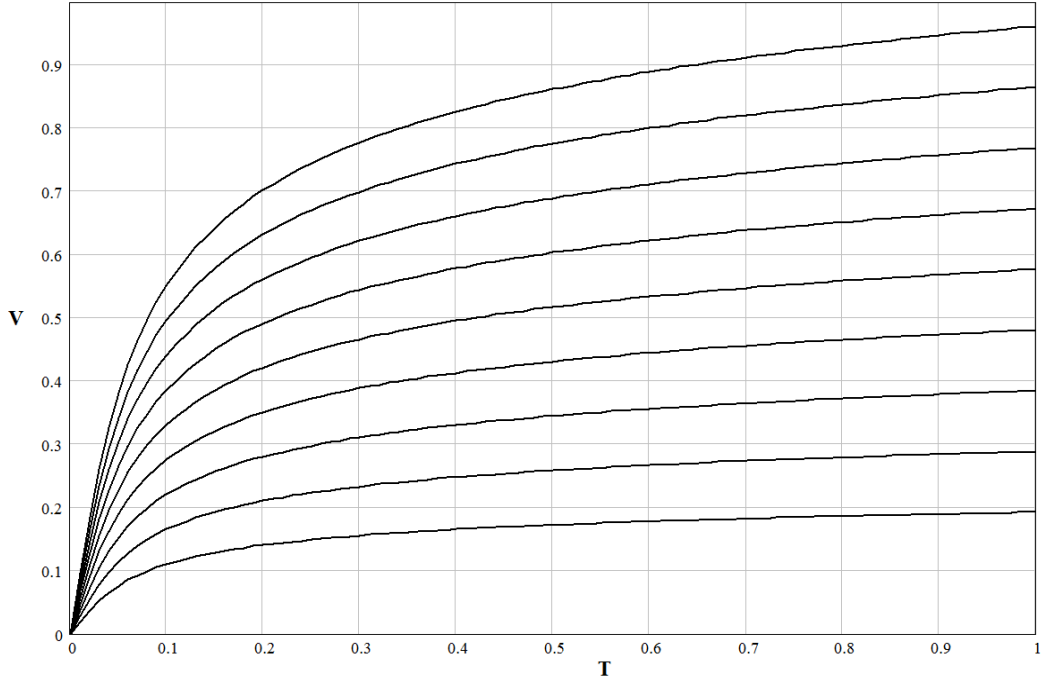

Fig. 2 Velocities of the cylindrical layers for the uniformly charged cylinder

This essential difference between the cylindrical and spherical areas is caused by the fact that the spherical area has the acceleration $\sim \dfrac{1}{R^2}$, that is enough for the limit of the wave velocity. In cylindrical area the acceleration $\sim \dfrac{1}{R}$, that is not enough for the limit of the wave velocity in 2D case. Arrangement of $\sim \dfrac{1}{R^2}$ or $\sim \dfrac{1}{R}$ is determined by the dimension of space in which the problem is under solution. For the cylindrical area the problem is actually solved for 2D space. From physical point of view, the fact that $V_{max}^c$ is limitless connects with the limits of a classical (non-relativistic) approximation.

Let us rewrite the expression (23) for the spherical case:

$$V_s(t) = V_{max}^s \sqrt{\dfrac{P_s(\lambda_s t)-1}{P_s(\lambda_s t)}}. \tag{24}$$

Let us express the function $P_s$ from (24) by $V_s$ and $V_{max}^s$:

$$P_s(\lambda_s t) = \dfrac{1}{1-\eta^2},\quad \eta = \dfrac{V_s}{V_{max}^s}. \tag{25}$$

Substituting (25) into (10) we obtain:

$$R_s(t) = \dfrac{R_0}{1-\left[\dfrac{V_s(t)}{V_{max}^s}\right]^2}. \tag{26}$$

It is noteworthy to say that the maximum probable velocity may be determined using the law of conservation of energy. The system is in the quiescent mode at initial time; that is why there is only the potential energy of the electric field of the charged system. Such potential energy may be determined as work of the electric field transferring the charge $q$ from the minimum distance $r_{min}$ to infinity:

$$U = A = \int_{r_{min}}^{+\infty} F dr = \int_{r_{min}}^{+\infty} \frac{qQ}{r^2} \frac{1}{4\pi\varepsilon_0} dr = -\frac{qQ}{4\pi\varepsilon_0} \frac{1}{r}\Big|_{r_{min}}^{+\infty} = \frac{qQ}{4\pi\varepsilon_0} \frac{1}{r_{min}}, \qquad (27)$$

where $Q$ is the charge in the sphere of radius $R_0$. Here we intentionally have used $r_{min} \neq R_0$ for the reason that is described further. At $t \to +\infty$ the whole potential energy is to become kinetic $T = \dfrac{m_q \left[V_{max}^s\right]^2}{2}$. Equating kinetic and potential energies we obtain the formula for the maximum velocity $V_{max}^s$ which is expressed by the initial minimum distance $r_{min}$

$$\frac{qQ}{4\pi\varepsilon_0} \frac{1}{r_{min}} = \frac{m_q \left[V_{max}^s\right]^2}{2} \Rightarrow V_{max}^s = \sqrt{\frac{qQ}{2\pi\varepsilon_0 m_q r_{min}}}. \qquad (28)$$

Let us consider the problem of the minimum probable distance $r_{min}$. There is a term of «classical electron radius» (CER) in classical electrodynamics. This value is known to have the following view:

$$r_0 = \frac{e^2}{4\pi\varepsilon_0 m_e c^2}, \qquad (29)$$

where $c$ − velocity of light in vacuum. Formula (29) is obtained due to equating the potential energy of the electric field electron to the electron rest energy $m_e c^2$.

If the dimension of the sphere $R_0$, which contains the charge $Q$, exceeds CER significantly, i.e. $R_0 \gg r_0$, then we can use $r_{min} = r_0$ from (29). Let us consider the case when $R_0 = r_0$ and $Q = q = e$, then it is logically to use $2r_0$ as the minimum distance $r_{min}$, as the minimum distance between the centers of two similar spheres is two radii of them. Substituting $r_{min} = 2r_0$ into (29) for the maximum velocity $V_{max}^s$, we obtain:

$$V_{max}^s = c. \qquad (30)$$

Substituting (30) into (26), we obtain:

$$R_s = \frac{R_0}{1 - \dfrac{V_s^2}{c^2}}. \qquad (31)$$

The expression (31) is obtained in suppositions of classic without taking relativity into account. Though, the existence of the maximum front velocity is correct. The sphere density is maximal at initial time. If the charge density is too big, the electric field created by them is big as well. If the electric field is big, the Coulomb force, affecting the charges, is maximum. Force is directly connected with acceleration according to the Newton's second law. That is why the charged particles have the maximum acceleration at initial time. Acceleration leads to the velocity increase. As the sphere becomes larger, the density decreases. That means that the Coulomb force decreases and acceleration aims to zero. As a result, velocity becomes a constant value $V_{max}^{s}$, as there is almost no acceleration left.

Let us illustrate the obtained results for the spherically symmetric initial distribution of the charge density $\rho_0(r)$.

### 1.1.1 Uniform initial distribution

Let density be $\rho_0(r) = const$. Let us determine $\lambda_s(r)$ at $\rho_s(R,0) = \rho_0 = const$, ($R\big|_{t=0} = rP_s(\lambda(r)t)\big|_{t=0} = r$ is taken into account):

$$\gamma_s(r) = \frac{Q(r)}{4\pi\varepsilon_0}\frac{q}{m} = \frac{\rho_0}{\varepsilon_0}\frac{q}{m}\int_0^r x^2 dx = \frac{\rho_0}{\varepsilon_0}\frac{\delta r^3}{3}, \qquad (32)$$

$$\lambda_s(r) = \frac{\sqrt{2\gamma_s(r)}}{r^{3/2}} = \sqrt{\frac{2}{3}\frac{\delta\rho_0}{\varepsilon_0}}. \qquad (33)$$

Expression (33) shows that $\lambda_s(r)$ does not depend on $r$, it is a constant value. Substituting (33) into (21), we obtain:

$$\rho_s(R,t) = \frac{\rho_0}{P_s^{3}\left(\sqrt{\frac{2}{3}\frac{\delta\rho_0}{\varepsilon_0}}t\right)}. \qquad (34)$$

The obtained (34) determines the evolution of the density function of a uniformly charged sphere at initial time. The formula (34) shows that the density inside the sphere stays uniform, i.e. it does not depend on the coordinate, it depends on time. Fig. 3 illustrates the evolution of the charge density (34).

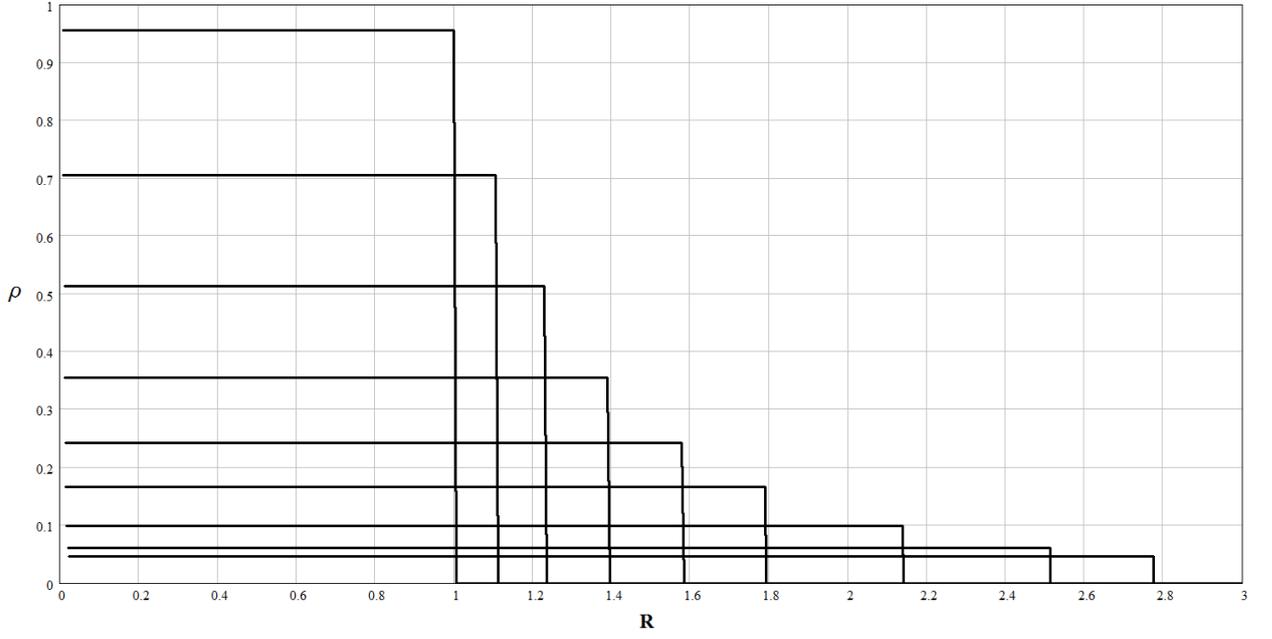

Fig. 3 Evolution of the charge density for the uniformly charged sphere

Fig. 4 presents the diagram of characteristics (10) for the uniformly charged sphere. Characteristics do not cross each other, and it corresponds with the supposition about non-crossing of the spherical layers of the charged sphere. The characteristics equation (10) may be rewritten without function $P$:

$$\sqrt{\frac{2}{3}\frac{\delta\rho_0}{\varepsilon_0}}t = \left(\sqrt{\left(\frac{\rho_0}{\rho_s}\right)^{1/3}\left(\left(\frac{\rho_0}{\rho_s}\right)^{1/3}-1\right)} + \operatorname{arcch}\left(\frac{\rho_0}{\rho_s}\right)^{1/6}\right) = F\left(\left(\frac{\rho_0}{\rho_s}\right)^{1/3}\right). \quad (35)$$

Expression (35) illustrates in an explicit form the dependence between time and density of the sphere charge.

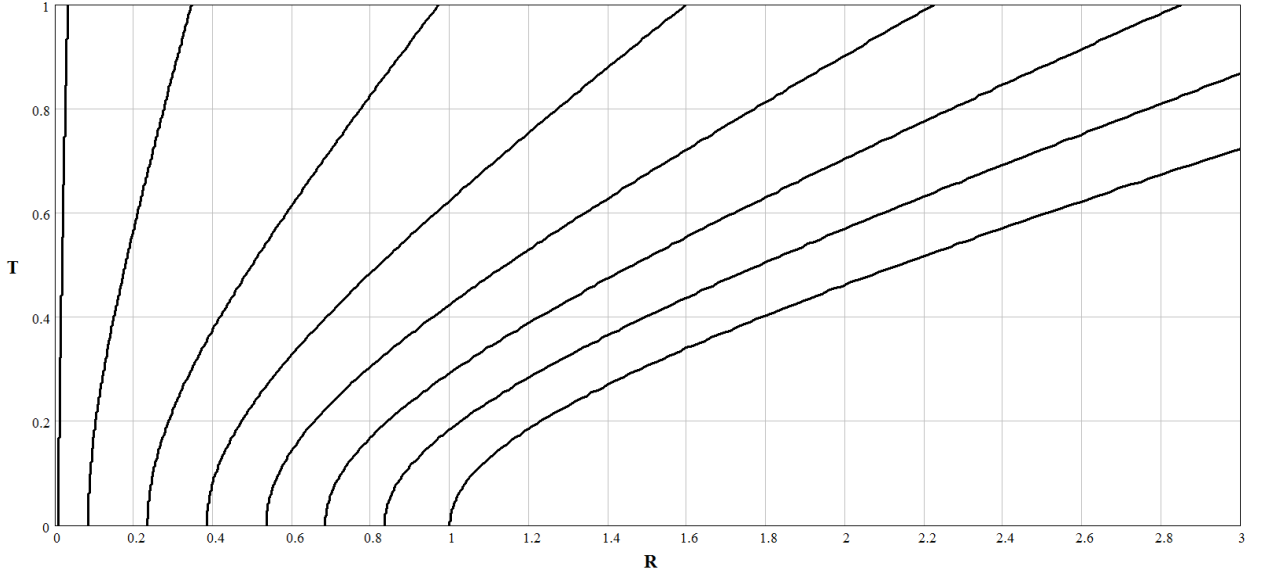

Fig. 4 Characteristics for the uniformly charged sphere

### 1.1.2 Lognormal initial distribution

Let us consider the charge probability density function using the following law:

$$\rho_0(r) = \frac{Q_{total}}{2\pi r^2} p_n(2r), \quad p_n(r) = \frac{1}{\sqrt{2\pi}\sigma r} e^{-\frac{(\ln(r)-\mu)^2}{2\sigma^2}}, \quad (36)$$

where $\mu$, $\sigma$, $Q_{total}$ are constant values. Let us determine $Q_s(R_0)$ integrating the expression (26) with the volume of a sphere of radius $R_0$:

$$Q_s(R_0) = 4\pi \int_0^{R_0} x^2 \frac{Q_{total}}{2\pi x^2} \frac{1}{\sqrt{2\pi}\sigma 2x} e^{-\frac{(\ln(2x)-\mu)^2}{2\sigma^2}} dx = \frac{Q_{total}}{2}\left(1+\mathrm{erf}\left(\frac{\ln(2R_0)-\mu}{\sigma\sqrt{2}}\right)\right), \quad (37)$$

where $\mathrm{erf}(x) = \frac{2}{\sqrt{\pi}} \int_0^x e^{-l^2} dl$. Using (37), we obtain the expression for $\lambda_s(r)$:

$$\lambda_s(r) = \frac{\sqrt{2\gamma_s(r)}}{r^{3/2}} = \sqrt{\frac{\delta Q_{total}}{4\pi\varepsilon_0 r^3}\left(1+\mathrm{erf}\left(\frac{\ln(2r)-\mu}{\sigma\sqrt{2}}\right)\right)}. \quad (38)$$

Substituting expressions (38) and (36) into (21), we obtain the final expression describing the evolution of the charge density function. The characteristics equation (10) takes the following form:

$$t = \frac{R_0^{3/2}}{\sqrt{\frac{\delta Q_{total}}{4\pi\varepsilon_0}\left(1+\mathrm{erf}\left(\frac{\ln(2R_0)-\mu}{\sigma\sqrt{2}}\right)\right)}}\left(\sqrt{\frac{R}{R_0}\left(\frac{R}{R_0}-1\right)}+\mathrm{arcch}\sqrt{\frac{R}{R_0}}\right). \quad (39)$$

Fig. 5 illustrates the diagrams of characteristics (39). Comparing Fig. 4 and Fig. 5 we can notice the principal difference in behavior of the characteristics. Fig. 5 shows that the characteristics cross each other. Fig. 6 presents the diagrams of the charge density in various moments of time. Crossing of the characteristics leads to the infinite increase of the charge density and to the probable occurrence of a shock wave.

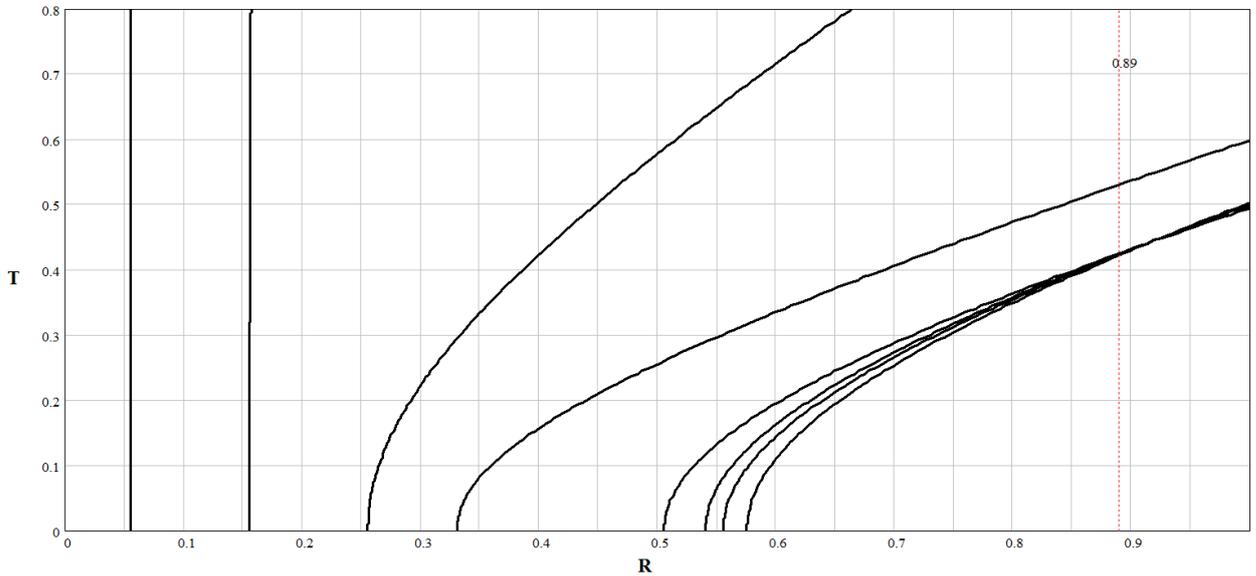

Fig. 5 Characteristics for the lognormal distribution

That is why solution (21) obtained in the supposition of non-crossing of the spherical layers (or by characteristics (39)) makes sense until the moment of occurrence of the characteristics crossing (see Fig. 5).

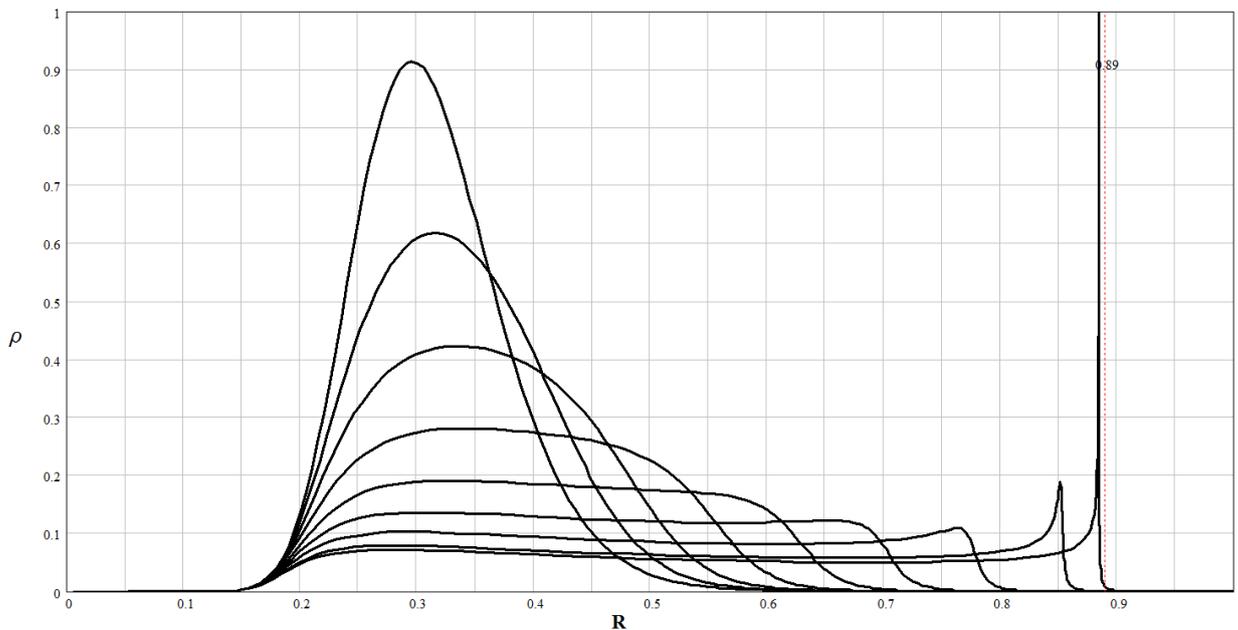

Fig. 6 Evolution of the density function for the lognormal distribution

It is noteworthy to say that similar diagrams (Fig. 3-6) are correct for the cylindrically symmetric distributions.

## 1.2 Gravitational interaction

Let us consider the system consisting of particles with the total mass $M$. Let the initial density of the mass $\rho_0(r)$ be known. Initial particle velocity is considered to be equal to zero in the system. We are going to solve the problem of evolution of the mass density function $\rho(r,t)$ of the given system. Due to the symmetry of the system, particles with the mass are influenced only by the gravitational attracting force directed along the radius. While solving the problem we are going to suppose that mass spherical (cylindrical) layers do not cross each other in the course of time. Let us write down an analogue of the Gauss theorem for gravitational field. Acceleration $\vec{g}$ is presented as the gravitational field intensity.

$$div\vec{G} = -\rho, \quad (40)$$

$$\vec{G} = \upsilon_0 \vec{g}, \quad \upsilon_0 = \frac{1}{4\pi k_g},$$

where $k_g$ is Newton's gravitational constant. Integrating (40) with the volume $V$ and surface $S$, and using the Gauss-Ostrogradsky Theorem, we obtain the Gauss theorem for gravitational field:

$$\int_S \vec{G}d\vec{s} = -\int_V \rho\, d\omega = -M, \quad (41)$$

where $M$ is mass in the volume $V$ which is limited by the surface $S$. In case of the spherically symmetric density distribution of the mass $\rho_s(r,t)$, vector $\vec{G}$ is perpendicular to the spherical surface $S$ that limiting certain mass $M$. As a result, formula (41) has the following form:

$$G_s(r,t) = \frac{1}{r^2}\int_0^r \rho_s(x,t)x^2 dx,$$

$$G_c(r,t) = \frac{1}{r}\int_0^r \rho_c(x,t) x\, dx. \quad (42)$$

Let us write down the Newton's second law for the mass $m$, which is on the surface of the sphere with radius $R$:

$$R_{tt} = -\frac{G_s(R,t)}{\upsilon_0} = -\frac{1}{\upsilon_0 R^2}\int_0^R \rho_s(x,t) x^2 dx,$$

$$R_{tt} = -\frac{G_c(R,t)}{\upsilon_0} = -\frac{1}{\upsilon_0 R}\int_0^R \rho_c(x,t)x\, dx. \quad (43)$$

Symbol «-» shows the movement to the origin of coordinates. As spherical (cylindrical) layers do not cross each other, the mass in the sphere (cylinder) with the radius $R_0 = R(t=0)$ stays constant during the whole movement. The value of the mass may be found using the following formula:

$$M_s(R_0) = 4\pi \int_0^{R_0} \rho_s(x,0) x^2 dx = 4\pi \int_0^{R(t)} \rho_s(x,t) x^2 dx,$$

$$M_c(R_0) = 2\pi \int_0^{R_0} \rho_c(x,0) x dx = 2\pi \int_0^{R(t)} \rho_c(x,t) x dx. \quad (44)$$

Taking (44) into account, expressions (43) will be rewritten as:

$$R_{tt} = -\frac{M_s(R_0)}{4\pi \upsilon_0} \frac{1}{R^2}, \quad R_{tt} = -\frac{M_c(R_0)}{2\pi \upsilon_0} \frac{1}{R}. \quad (45)$$

As a result we obtain the Cauchy problem for $R(t)$:

$$\begin{cases} R'' = -\gamma_s \dfrac{1}{R^2}, \\ R\big|_{t=0} = R_0, \\ R'\big|_{t=0} = 0, \end{cases} \qquad \begin{cases} R'' = -\gamma_c \dfrac{1}{R}, \\ R\big|_{t=0} = R_0, \\ R'\big|_{t=0} = 0, \end{cases} \quad (46)$$

where $\gamma_s = \dfrac{M_s(R_0)}{4\pi \upsilon_0}$, $\gamma_c = \dfrac{M_c(R_0)}{2\pi \upsilon_0}$ are constant values. The solution of problems (40) is determined in similar way as for the case of the electric field. As a result, we obtain the characteristics equations (10) where function $F$ and $\lambda$ have the following form:

$$F_s(x) = \sqrt{x(1-x)} + \arccos\sqrt{x}, \quad F_c(x) = \frac{\sqrt{\pi}}{2} erf\left(\sqrt{\ln\frac{1}{x}}\right), \quad (47)$$

$$\lambda_s(R_0) = \frac{\sqrt{2\gamma_s(R_0)}}{R_0^{3/2}}, \quad \lambda_c(R) = \frac{\sqrt{2\gamma_c(R)}}{2R}.$$

To determine evolution of the density function on the analogy of the electric field, we are going to evaluate the value $\dfrac{\Delta M}{\Delta R}$. $\Delta M$ is mass between two spherical (cylindrical) layers with radii $r$ and $r + \Delta r$ respectively. As a result:

$$\rho_s(R,t) = \frac{1}{P_s^2(\lambda_s(r)t)} \frac{\rho_0(r)}{\left[P_s(\lambda_s(r)t) - t\sqrt{\dfrac{1-P_s(\lambda_s(r)t)}{P_s(\lambda_s(r)t)}} \left[\dfrac{\rho_0(r)}{\upsilon_0 \lambda_s(r)} - \dfrac{3}{2}\lambda_s(r)\right]\right]},$$

$$\rho_c(R,t) = \frac{1}{P_c(\lambda_c(r)t)} \frac{\rho_0(r)}{\left[P_c(\lambda_c(r)t) - 2t\sqrt{\ln\dfrac{1}{P_c(\lambda_c(r)t)}} \left[\dfrac{\rho_0(r)}{4\upsilon_0 \lambda_c(r)} - \lambda_c(r)\right]\right]},$$

(48)

where $\rho(R,t) = \rho(rP(\lambda(r)t),t)$. Formula (48) determines the evolution of the mass density function with initial distribution $\rho_0(r)$ at the condition that the spherical (cylindrical) layers do not cross each other.

Analogous correlations (22) are correct for the velocity:

$$\frac{d}{dt}R_s(t) = V_s(t) = -R_0\lambda_s\sqrt{\frac{1-P_s(\lambda_s t)}{P_s(\lambda_s t)}},$$

$$\frac{d}{dt}R_c(t) = V_c(t) = -R_0\lambda_c 2\sqrt{\ln\frac{1}{P_c(\lambda_c t)}}. \qquad (49)$$

Let us consider special cases of initial distributions of the mass density $\rho_0(r)$ and obtain the evolution of the density function $\rho_s(r,t)$.

### 1.2.1 Uniform initial distribution

Let the density be $\rho_0 = const$. Let us determine $\lambda_s(r)$ at $\rho_s(r,0) = \rho_0 = const$ ($R|_{t=0} = rP_s(\lambda_s(r)t)|_{t=0} = r$ is taken into account), we obtain:

$$\gamma_s(r) = \frac{M(r)}{4\pi \upsilon_0} = \frac{\rho_0}{\upsilon_0}\int_0^r x^2 dx = \frac{\rho_0}{\upsilon_0}\frac{r^3}{3},$$

$$\lambda_s(r) = \frac{\sqrt{2\gamma_s(r)}}{r^{3/2}} = \sqrt{\frac{2}{3}\frac{\rho_0}{\upsilon_0}}. \qquad (50)$$

Expression (50) shows that $\lambda_s(r)$ does not depend on $r$, it is a constant value. Substituting (50) into (48), we obtain:

$$\rho_s(r,t) = \frac{\rho_0(r)}{P_s^3\left(\sqrt{\frac{2}{3}\frac{\rho_0}{\upsilon_0}}t\right)}. \qquad (51)$$

The obtained expression (51) determines the evolution of the density function of the sphere mass with initial distribution $\rho_0 = const$. As we can see from (51), the mass density inside the sphere stays uniform, it does not depend on the coordinate, it depends on time. Fig. 7 illustrates the evolution of the mass density (51).

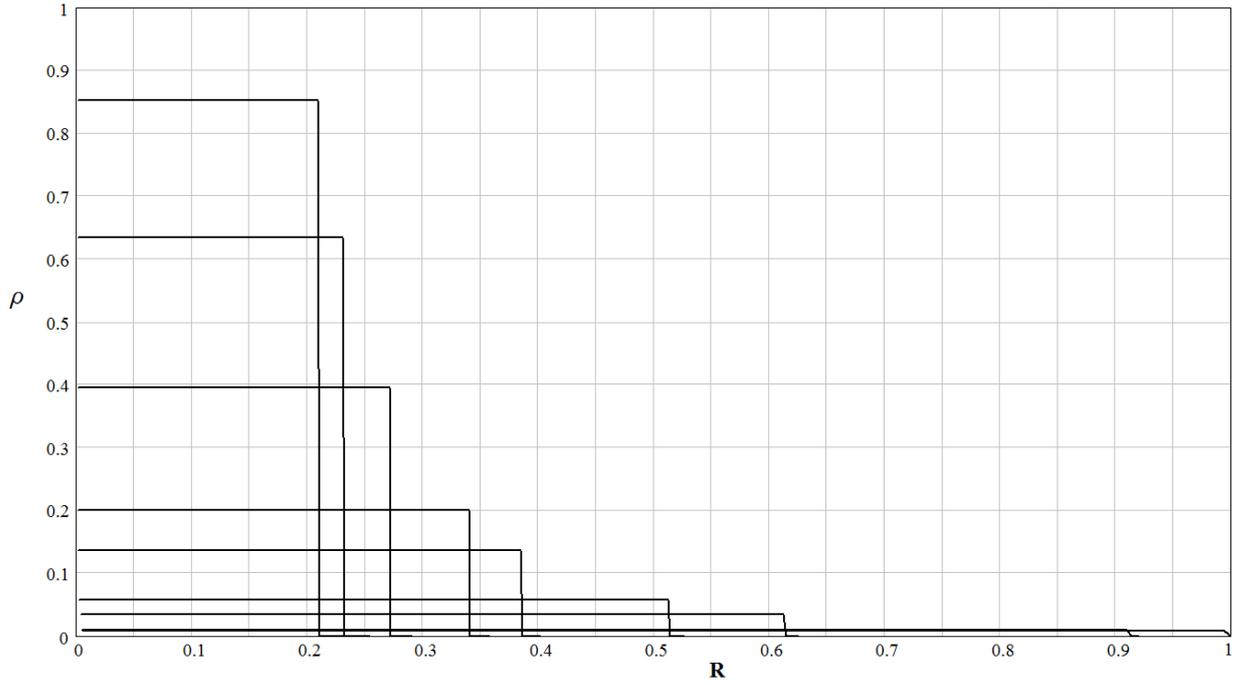

Fig. 7 Evolution of the density function for a uniform sphere

Diagrams of characteristics for a uniform sphere are presented in Fig. 8. Fig. 8 illustrates that there is a certain moment of time $T_0 = \dfrac{\pi}{2}\sqrt{\dfrac{3\upsilon_0}{2\rho_0}}$, when all the layers are in the center of a sphere. The layers do not cross each other up to the moment $T_0$, and solution makes sense.

The characteristics equation may be rewritten without function $P$:

$$\sqrt{\dfrac{2}{3}\dfrac{\rho_0}{\upsilon_0}}\, t = \left( \sqrt{\left(\dfrac{\rho_0}{\rho_s}\right)^{1/3}\left(1-\left(\dfrac{\rho_0}{\rho_s}\right)^{1/3}\right)} + \arccos\left(\dfrac{\rho_0}{\rho_s}\right)^{1/6} \right) = F\left(\left(\dfrac{\rho_0}{\rho_s}\right)^{1/3}\right). \quad (52)$$

Expression (52) illustrates in an explicit form the dependence between time and density of the sphere mass.

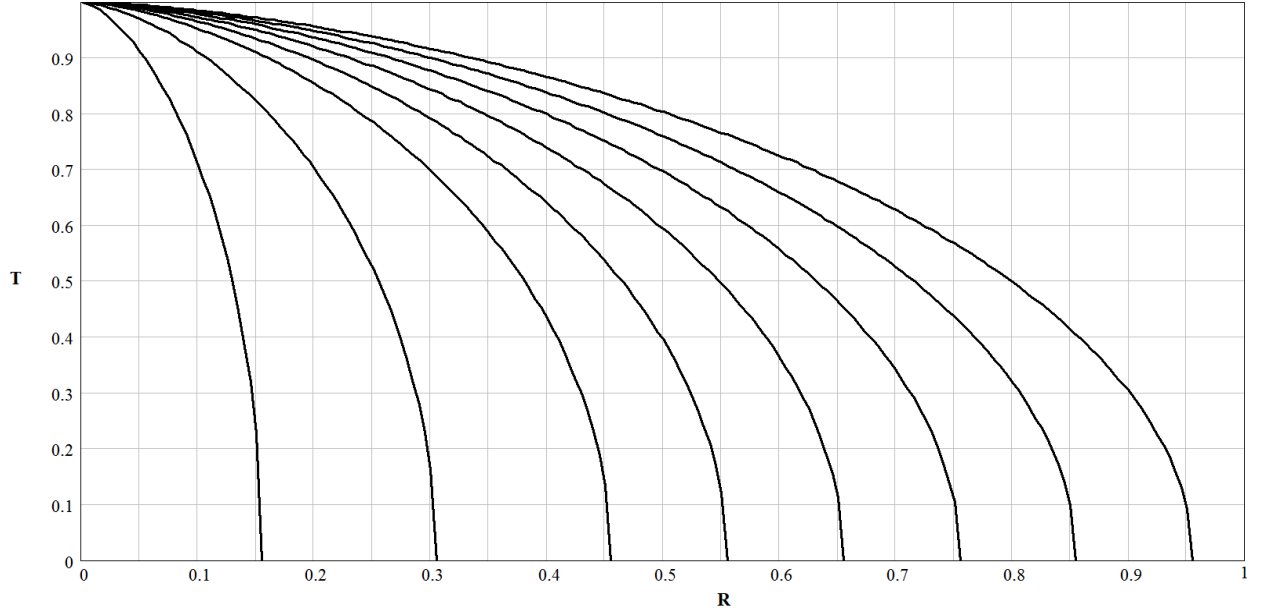

Fig. 8 Characteristics for a uniform sphere

**1.2.2 Lognormal initial distribution**

Let us consider the mass density distribution inside the sphere using the following law:

$$\rho_0(r) = \frac{M_{total}}{2\pi r^2} \rho_n(2r), \quad \rho_n(r) = \frac{1}{\sqrt{2\pi}\sigma r} e^{-\frac{(\ln(r)-\mu)^2}{2\sigma^2}}, \quad (53)$$

where $\mu$, $\sigma$, $M_{total}$ is constant values. Let us determine $M_s(R_0)$ by integrating expression (53) with the sphere volume of the radius $R_0$:

$$M_s(R_0) = 4\pi \int_0^{R_0} x^2 \frac{M_{total}}{2\pi x^2} \frac{1}{\sqrt{2\pi}\sigma 2x} e^{-\frac{(\ln(2x)-\mu)^2}{2\sigma^2}} dx = \frac{M_{total}}{2}\left(1 + erf\left(\frac{\ln(2R_0)-\mu}{\sigma\sqrt{2}}\right)\right), \quad (54)$$

where $erf(x) = \frac{2}{\sqrt{\pi}} \int_0^x e^{-l^2} dl$. Using (54), we obtain the expression for $\lambda_s(r)$:

$$\lambda_s(r) = \frac{\sqrt{2\gamma_s(r)}}{r^{3/2}} = \sqrt{\frac{M_{total}}{4\pi \upsilon_0 r^3}\left(1 + erf\left(\frac{\ln(2r)-\mu}{\sigma\sqrt{2}}\right)\right)}. \quad (55)$$

Substituting (55) and (53) into (48), we obtain the final expression describing the evolution of the mass density function for the case of the initial density distribution in the form of the lognormal distribution. The characteristics equation takes the following form:

$$t = \frac{R_0^{3/2}}{\sqrt{\frac{M_{total}}{4\pi\upsilon_0}\left(1 + erf\left(\frac{\ln(2R_0)-\mu}{\sigma\sqrt{2}}\right)\right)}} \left(\sqrt{\frac{R}{R_0}\left(1 - \frac{R}{R_0}\right)} + \arccos\sqrt{\frac{R}{R_0}}\right) \quad (56)$$

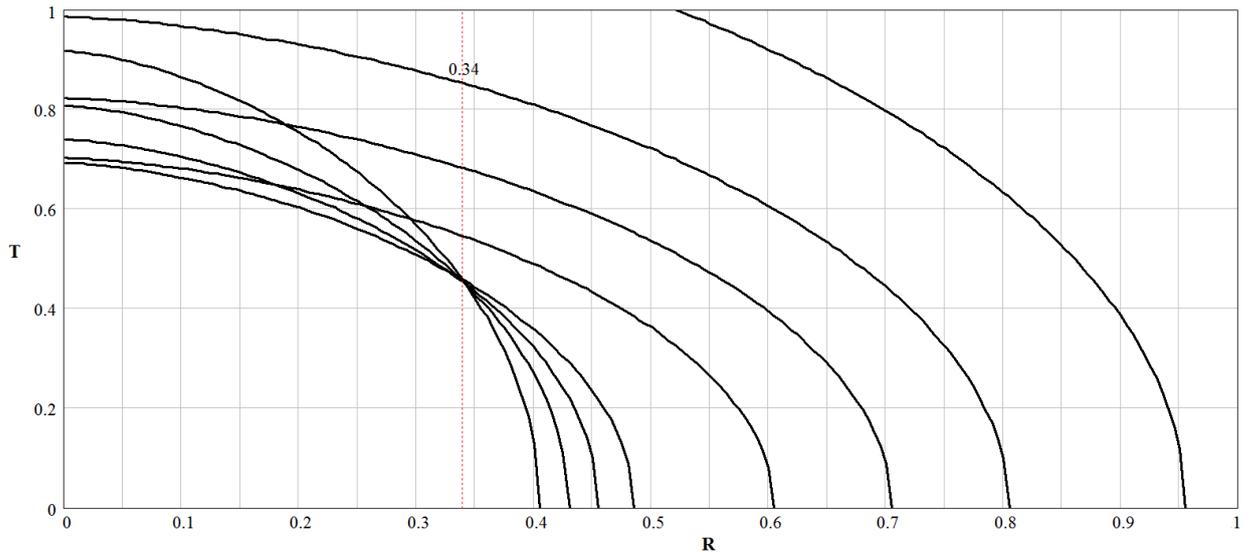

Fig. 9 Characteristics for the lognormal distribution

Fig. 9 illustrates the diagrams of characteristics (56). Comparing Fig. 8 and Fig. 9 we can see different character of the characteristics behavior. In Fig. 9 the characteristics cross each other in the certain radius, not in the center of a sphere.

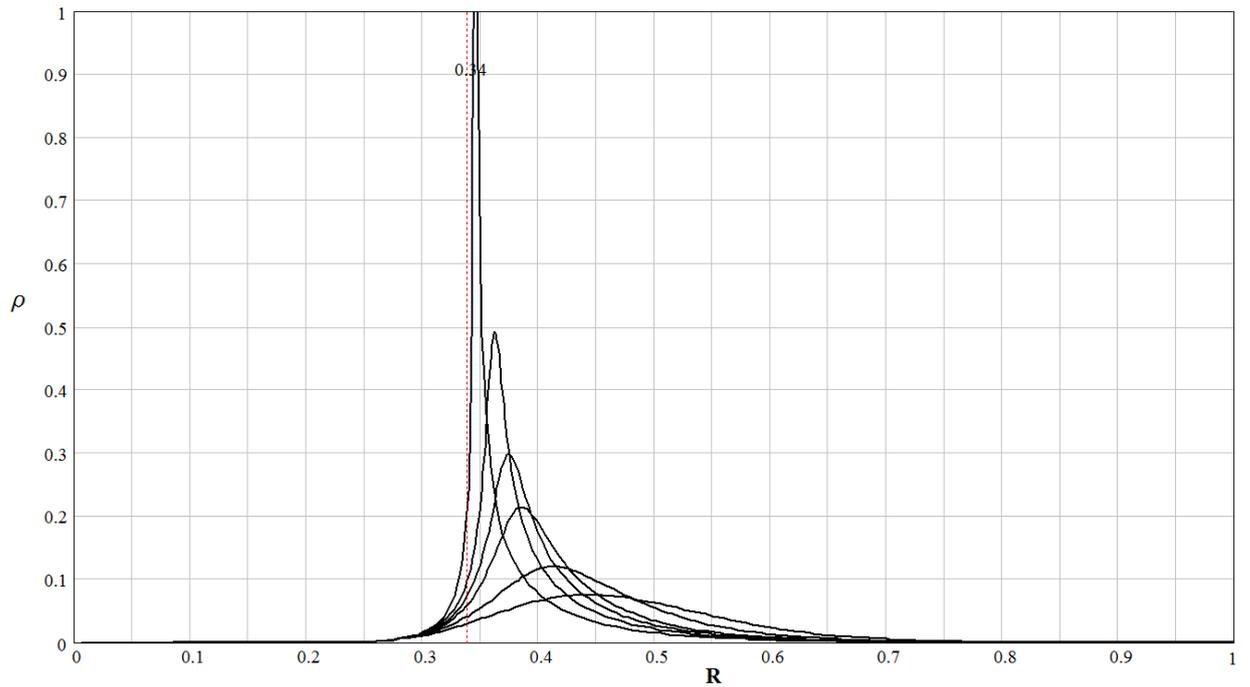

Fig. 10 Evolution of the density function for the lognormal distribution

Fig. 10 illustrates the diagrams of the mass density at various moments of time. Crossing of the characteristics leads to the infinite increase of the mass density on a certain spherical layer.
That is why solution (48), obtained in the supposition on non-crossing of the spherical layers, makes sense up to the occurrence of the characteristics crossing (see Fig. 9).

**2. Solution of the shock wave type for the Schrödinger equation**

The charge/mass density functions (21), (48) and velocity functions (21), (49) corresponding with them and obtained in the Sect.1, as it was noted before [1], are the solutions of equations (i.2). In addition the density functions $\rho(r,t)$ of the charge/mass acquire the sense of the probability density functions $f(r,t)$. Functions of the charge $Q(r,t)$ and mass $M(r,t)$ acquire the sense of the probability distribution functions $F(r,t)$ in this case. That is why it is necessary to use the value of the total charge/mass $Q_{total}$ / $M_{total}$ equal to 1, that corresponds with the normality condition of the wave function $\Psi(r,t)$.

Flow velocities of the charge/mass have the interpretation of the flow velocity of probability $\langle\vec{v}\rangle(r,t)$ in (i.2) equation. Velocities $V(t)$ correspond with concentric layers, the mechanical trajectory of which is described by the characteristics equations (10). Knowing the velocity $V(t)$ and the characteristics equation $R(t)$, we can determine the flow velocity of probability $\langle\vec{v}\rangle(r,t)$. For example, for the spherically symmetric distribution with the electric interaction (10), (22):

$$r(R_0,t) = R_0 P(\lambda(R_0)t) \stackrel{\text{det}}{\Rightarrow} G(r,t) = R_0,$$
$$u(R_0,t) \stackrel{\text{det}}{=} R_0 \lambda_s(R_0) \sqrt{\frac{P_s(\lambda_s(R_0)t)-1}{P_s(\lambda_s(R_0)t)}}, \quad (57)$$
$$\langle\vec{v}\rangle(r,t) = u(G(r,t),t)\vec{e}_r.$$

As the flow of charge is laminar in this example, statement (i.1) does not have the vortical component of the velocity $\vec{A}$. Therefore, according to (i.1), (i.5) the following statement in correct for $\langle\vec{v}\rangle(r,t)$:

$$\langle\vec{v}\rangle(r,t) = -\alpha\nabla\Phi(r,t) = -2\alpha\nabla\varphi(r,t). \quad (58)$$

Taking (57) into account and integrating (58) with the radius, we can determine the phase $\varphi(r,t)$ of the wave function, which is scalar potential of the velocity of the probability flow $\langle\vec{v}\rangle(r,t)$. Fig. 11 illustrates the obtained (using (57)) distribution of the velocity of the probability flow $\langle\vec{v}\rangle(r,t)$ along the radius inside the sphere at various moments of time. The initial distribution of the probability density $f_0(r) = \rho_0(r)$ and corresponds with the case of the shock wave (36). It is seen that the front becomes vertical in the course of time, and it is typical of a shock wave.

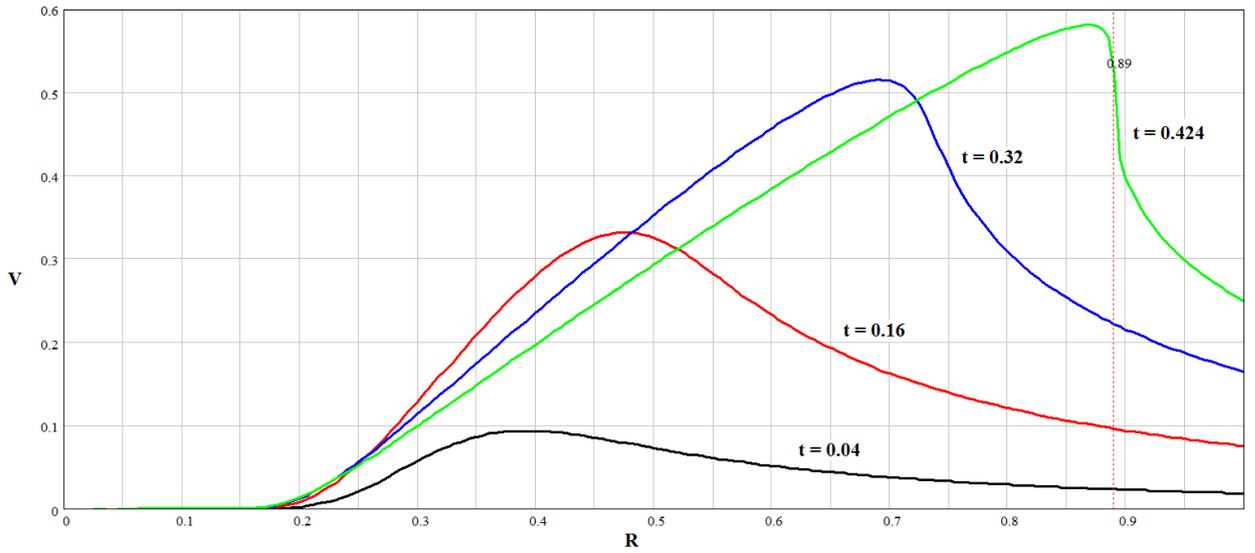

Fig. 11 Velocity of the probability flow

Fig. 12 illustrates the corresponding distribution of the wave function phase $\varphi(r,t)$ or within the accuracy of the coefficient of the velocity scalar potential $\langle \vec{v} \rangle (r,t)$.

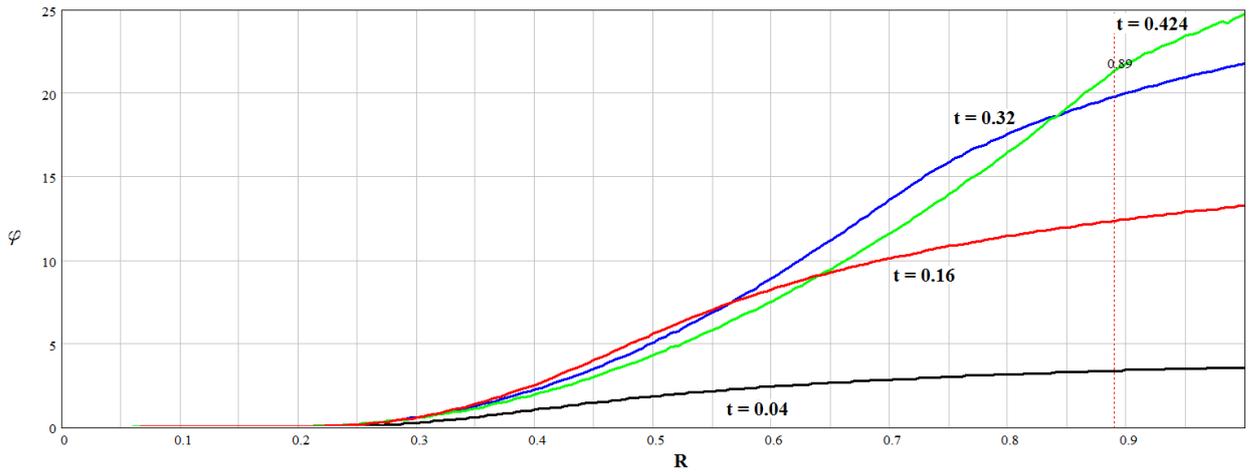

Fig. 12 Scalar potential of the velocity of the probability flow

Knowing the phase $\varphi(r,t)$ and the probability density function $f(r,t) = \rho(r,t)$, we can determine the wave function $\Psi(r,t)$ itself $\Psi(r,t) = \sqrt{f(r,t)} e^{i\varphi(r,t)}$. Fig. 13 illustrates the diagram of the wave function up to the occurrence of a shock wave. Function $\Psi(r,t)$ is not univalent.

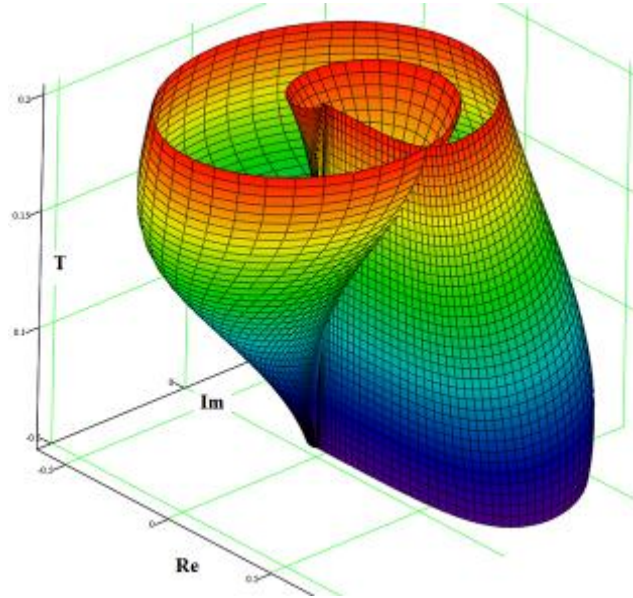

Fig. 13 Evolution of the wave function

Using (4) we can determine the potential $U(r,t)$ that is included into the Schrödinger equation (i.3). Fig. 14 illustrates the distribution $U(r,t)$ along the radius at various moments of time.

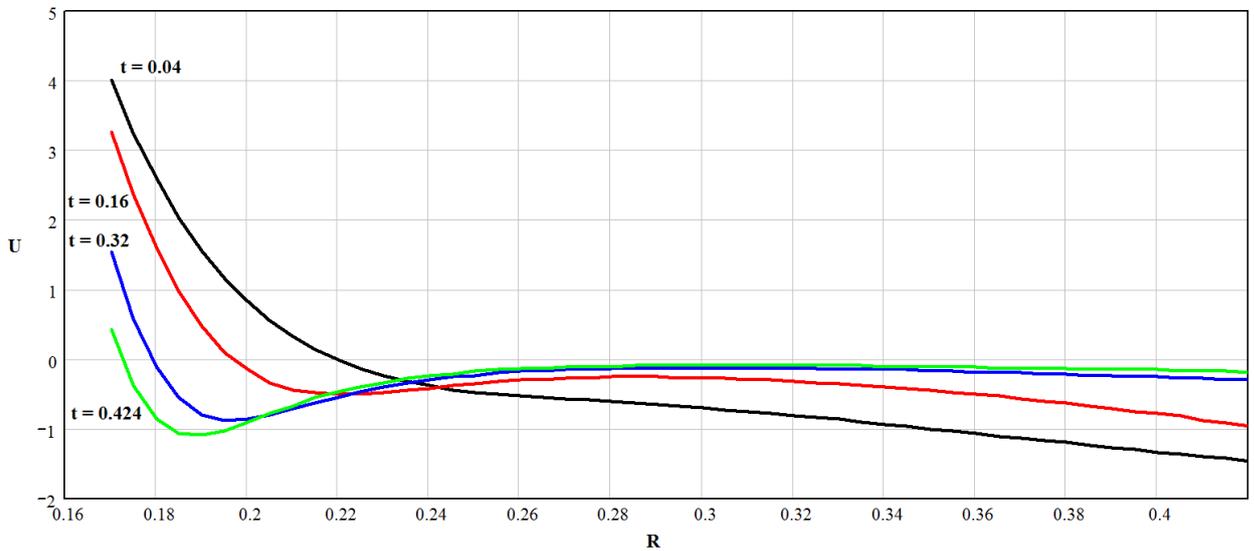

Fig. 14 Evolution of the potential

## 3. Interpretation of the deterministic and probabilistic description of a system

Fig. 15, 16 illustrate the probability distribution functions $F(r,t)$ for two cases of the initial density of the probability distribution $f_0(r)$. Fig. 15 corresponds with the initial distribution $f_0(r) = const$ without a shock wave. Fig. 16 corresponds with the initial distribution (36) with a shock wave. Fig. 15, 16 illustrate the evolution of the probability distribution

function in the course of time. Both figures show the diagrams of characteristics (10) marked with red color corresponding with various concentric layers.

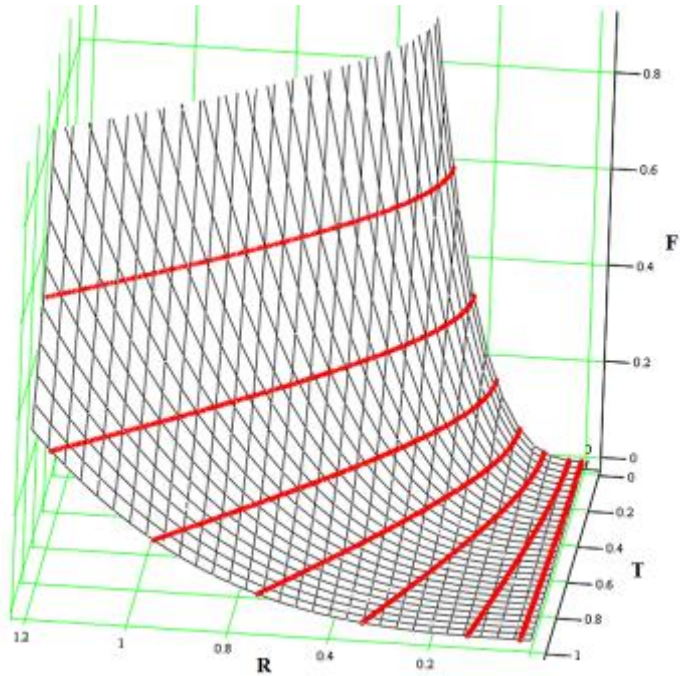

Fig. 15 The probability distribution function without a shock wave

It is noteworthy to say that curves, corresponding with the characteristics, completely belong to the surface of the probability distribution function $F(r,t)$. Thus, the surface of the probability distribution function fully consists of such curves [12].

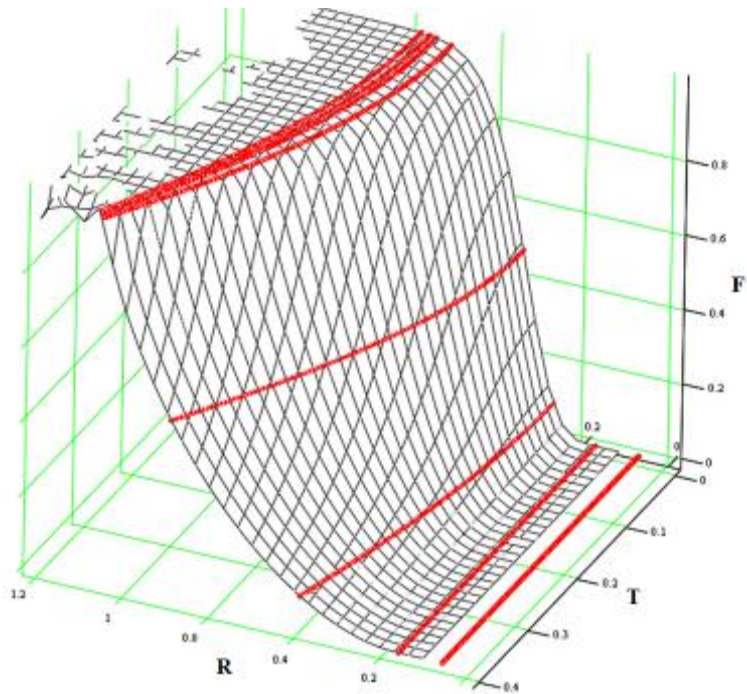

Fig. 16 The probability distribution function with a shock wave

From physical point of view, characteristics equations are classical equations of the concentric layers movement. From geometric point of view, the surface of the probability distribution function is developed by these curves. It turns out that on the one hand there is a deterministic description of the evolution of the charge/mass density using classical equations of motion. On the other hand, the probabilistic description is used through the probability distribution function. However, both cases include the same surface which can be interpreted both as the probability distribution function and as the set of classical mechanical trajectories (characteristics).

It is noteworthy to say that the probability distribution function $F(r,t)$ (or of the charge/mass) agrees with the following initial-boundary problem:

$$\begin{cases} \dfrac{\partial F(r,t)}{\partial t} + v(r,t)\dfrac{\partial F(r,t)}{\partial r} = 0, \\ \dfrac{\partial v(r,t)}{\partial t} + v(r,t)\dfrac{\partial v(r,t)}{\partial r} = \kappa \cdot S(r) \cdot F(r,t), \\ v(r,t)\big|_{t=0} = 0, \\ F(r,t)\big|_{t=0} = F_0(r), \end{cases} \qquad (59)$$

where

$$S(r) = \begin{cases} \dfrac{1}{r^2}, \text{ sphere,} \\ \dfrac{1}{r}, \text{ infinte cylinder,} \end{cases} \quad F(r,t) = \begin{cases} 4\pi \displaystyle\int_0^r x^2 f(x,t)\,dx, \text{ sphere,} \\ 2\pi \displaystyle\int_0^r x f(x,t)\,dx, \text{ infinite cylinder.} \end{cases} \qquad (60)$$

The constant value $\kappa$ adopts the following according to the type of physical problem:

$$\kappa = \begin{cases} \dfrac{\delta}{\varepsilon_0}, \quad \text{electric field,} \\ -\dfrac{1}{\upsilon_0} = -4\pi k_g, \text{ gravity field,} \\ \dfrac{\delta}{\varepsilon_0} - \dfrac{1}{\upsilon_0} = \dfrac{1}{\varepsilon_0}(\delta - \delta_0),\ \delta_0 = \dfrac{\varepsilon_0}{\upsilon_0}, \text{ electric and gravity fields.} \end{cases}$$

The first equation (59) corresponds to the continuum equation $f_t + \text{div}(\vec{v} f) = 0$ and the second one in (59) corresponds to the motion equation (4), (45) in the Euler coordinates. Consequently the solution obtained (21) and (48) are the problem (59) solutions taking in to account the expressions (3) and (44) correspondingly.

**Conclusion**

The results obtained:
1. In the paper the idea to obtain the solution in the quantum case which is known in the classical case as «shock wave» was realized. The base of the same consideration was the link between the classical continuity equation for the density distribution function and the quantum Schrödinger one for the wave function [1]. This solution in the paper was obtained (the quantum «shock wave» solution) (21), (48).

2. The exact solutions of the hydrodynamic equations (59) for the charge density and mass one was obtained in the electric (21) and gravitate (48) self-consistent field for the spherical and cylinder symmetry. In the case when the initial density distribution is inhomogeneous it is possible the «shock wave» existence. It appears mathematically as the cross point on the characteristics graph (see Fig. 5,9). Particularly, in the paper on the example of the inhomogeneous lognormal distribution it was considered (Section 1.1.2 and Section 1.2.2).
3. The common geometrical properties of the solutions in the classical and quantum case were investigated (see Fig. 15, 16). It is obtained that the classical motion trajectories for the concentric layers of the charge density distribution or the mass one are the forming the probability distribution function surface. These classical motion trajectories are characteristic equations solutions (see Fig. 15, 16). In the another hand that is from the probability point of view there is the probability distribution function which the probability distribution describes. By the way the distribution function surface consists of the classical motion trajectories. In the other words the surface consists of the characteristics solutions. That is mathematically from the geometrical point of view we have the object, which we can to interpret on the two ways. One the way is to interpret as the surface. And another one is to interpret it as the set of the curves (characteristics) forming this surface. In the case when the surface is considering we deal with the probability (quantum) approach. In the case when we consider the surface as characteristics set the classical continuum mechanics take a place (deterministic approach).